\documentclass{article}
\usepackage{spconf,amsmath,graphicx, adjustbox, array}
\usepackage{amsfonts, dsfont}
\usepackage{diagbox}
\usepackage[table]{xcolor}

\usepackage{hyperref}

\title{EXPLORING SELF-SUPERVISED CONTRASTIVE LEARNING OF SPATIAL SOUND EVENT REPRESENTATION}
\name{Xilin Jiang, Cong Han, Yinghao Aaron Li, Nima Mesgarani}
\address{Department of Electrical Engineering, Columbia University, USA}

\begin{document}
\ninept
\maketitle
\begin{abstract}
In this study, we present a simple multi-channel framework for contrastive learning (MC-SimCLR) to encode `what' and `where' of spatial audios. MC-SimCLR learns joint spectral and spatial representations from unlabeled spatial audios, thereby enhancing both event classification and sound localization in downstream tasks. At its core, we propose a multi-level data augmentation pipeline that augments different levels of audio features, including waveforms, Mel spectrograms, and generalized cross-correlation (GCC) features. In addition, we introduce simple yet effective channel-wise augmentation methods to randomly swap the order of the microphones and mask Mel and GCC channels. By using these augmentations, we find that linear layers on top of the learned representation significantly outperform supervised models in terms of both event classification accuracy and localization error. We also perform a comprehensive analysis of the effect of each augmentation method and a comparison of the fine-tuning performance using different amounts of labeled data.
\end{abstract}
\section{Introduction}
\label{sec:intro}
The majority of audio pre-training models are centered on learning robust auditory representations, facilitating the identification of `what' the sound source is \cite{SSAST, huang2022masked, chen2022beats}. However, a complete representation of audio that can be used in a broader range of applications needs to include spatial attributes, as location is an intrinsic feature of all sound objects.  In many applications, including acoustic surveillance, environmental monitoring, augmented reality, and autonomous vehicles, where ambient intelligence and acoustic awareness are desired, it is not sufficient to merely classify what and when sound events happen, but we also need to locate them in space. Learning disjoint representation of spectral and spatial properties leads to unnecessary problems, such as linking each sound event with its location. Moreover, 
a common real-world challenge revolves around the absence of annotations for either spectral or spatial attributes in audio data, rendering large-scale supervised training unfeasible.
To tackle both issues, we propose \textit{a simple multi-channel framework for contrastive learning} (MC-SimCLR), the first self-supervised representation learning framework of multi-channel audio, to jointly learn spectral and spatial attributes of audios without supervision.

MC-SimCLR is an adaptation of \textit{a simple framework for contrastive learning} (SimCLR) \cite{simclr} for unlabeled multi-channel audio data. The core of our framework is Multi-level Data Augmentation, a chain of augmentation applying to the waveform, Mel spectograms and generalized cross-correlation (GCC) features. We adopt existing augmentations that operate on two-dimensional features and also introduce new argmentations that operate on the channel dimension. Specifically, we randomly swap the order of the microphones to generate more training samples and drop entire channels of features to discourage overfitting on specific channels. We assess the efficacy of the framework with the task of sound event localization and detection (SELD). The experimental results show that using MC-SimCLR embedding leads to improved event classification accuracy and reduced azimuth prediction error compared to training from scratch. This clearly underscores MC-SimCLR's proficiency in extracting both spectral and spatial-discriminative features from unlabeled multi-channel audio data.

\begin{figure*}[t]
  \centering  \includegraphics[width=0.95\textwidth]{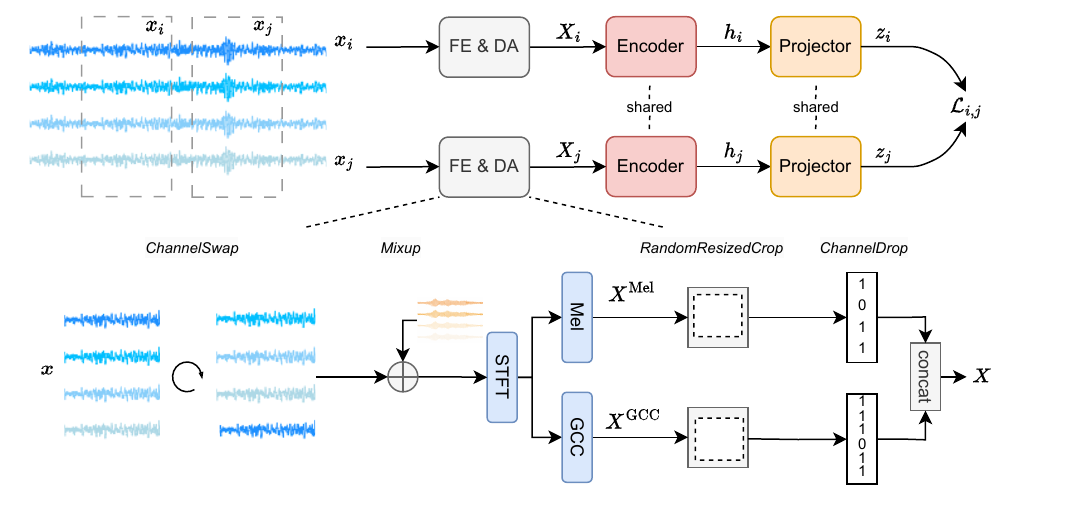}
  \caption{An overview of MC-SimCLR (top) with an in-depth visualization of feature extraction and data augmentation (FE \& DA) module (bottom) to showcase where and in which order each augmentation is applied.}
  \label{fig:overview}
\end{figure*}

\section{Related Works}
\label{sec:works}
Supervised training for SELD has witnessed significant progress in recent years \cite{salsa_lite, seld_features, crnn, seld_mhsa, four-stage, mcs}. However, a challenge in real-world scenarios is the lack of labels pertaining to either spectral or spatial attributes in audio data, which makes large-scale supervised training impracticable. Self-supervised learning has been a promising representation learning approach without the necessity for explicit labels. The learned representations often serve as input features for downstream tasks, diminishing the demand for extensive labeled training data but improving task performance. Past research has studied self-supervised learning approaches for sound event detection (SED) and sound source localization (SSL) separately. 

In the field of SED, contrastive learning frameworks based on SimCLR \cite{simclr} maximize the similarity of segments from the same recording and minimize the similarity of segments from different recordings \cite{cl_sound, cl_general_audio, wang2021multiformat}. This objective discriminates segments of different classes, as segments from the same recording are more likely to share the same label than those from different recordings. Self-distillation method \cite{byol} achieves the same goal without contrasting multiple segments by having an online encoder predict the embedding of the target encoder \cite{byol_audio}. Additionally, transformer patch modeling method also learns discriminative features for event classification through a proxy task of predicting and reconstructing masked spectrogram patches from unmasked ones \cite{SSAST}.

Self-supervised learning for SSL has been less studied. Although traditional signal processing techniques such as the Delay-and-Sum Beamformer and SRP-PHAT \cite{srp-phat} do not require supervised training, their performance degrades significantly in the presence of noise and reverberation. Recently, contrastive random work \cite{time-delay} has been utilized to estimate interaural time difference (ITD) from the learned embedding of each channel.  However, this set of embeddings is not class-discriminative for the SED task, and the variable number (which is not fixed to one) of embeddings also imposes limitations on its applicability. In another study \cite{francl2022modeling}, direction-variant features are extracted from binaural recordings through contrastive learning. However, their framework thresholds on known head rotation to define positive or negative samples in terms of source direction, and the resulting embedding is primarily intended for localization purpose as well.

Several other efforts have incorporated self-supervised pre-trained models on single-channel spectrograms for multi-channel SELD task \cite{pretrain_ssast, pretrain_ssast2}. These approaches necessitate additional components and training procedures to establish connections between event categories and their spatial locations, leading to increased computation and complexity. In our self-supervised learning framework, we learn a single embedding that serves the dual purpose of classification and localization, significantly reducing the required efforts for integration and fine-tuning.

\section{MC-SimCLR}
\label{sec:contrastive}
MC-SimCLR follows the methodology of single-channel SimCLR by training a model to differentiate between similar and dissimilar data points. For SELD, `similar' data points refers to audio samples that share both identical class labels and nearby locations. Conversely, `dissimilar' samples can vary in either of these aspects. To effectively capture this similarity from unlabeled multi-channel audios, MC-SimCLR extracts both spectral and spatial features, applies augmentation on various features and dimensions, and optimizes a contrastive objective during training. Fig. \ref{fig:overview} depicts the entire framework with an emphasis on feature extraction and data augmentation stage.

\subsection{Input Features} \label{features}
We follow the common assumption in audio contrastive learning that two random patches cropped from the same recording are considered as the positives (of the same class) \cite{cl_sound, cl_general_audio, wang2021multiformat, cl_music}. We further assume that the source remains stationary in the recording, ensuring that two patches also share the same spatial location. We crop two random patches, denoted as $x_i, x_j \in \mathbb{R}^{M\times N}$, from the same waveform to form a pair of positive samples. Each of these patches is a waveform with $M$ channels and a duration of one second. Shorter recordings are zero-padded to one second, and there can be an overlap between two random patches from the same recording.

Next, we extract Mel spectrograms and Generalized Cross-Correlation Phase Transform (GCC) features from the wave patches, which are common in SELD literature \cite{dcase, four-stage, two-stage}. Specifically, we extract the log-Mel spectrogram of audio for each microphone, $X^{\text{Mel}} \in \mathbb{R}^{M \times F \times T}$, and GCC features for each pair of microphones, $X^{\text{GCC}} \in \mathbb{R}^{{M \choose 2} \times F \times T}$, defined as
\begin{align} \label{gcc}
    X^{\text{GCC}}_{i, j}(\tau, t) = \mathcal{F}^{-1}_{f \rightarrow \tau} \frac{X_i(f, t) X_j^{\ast}(f, t)} {|X_i(f, t)| |X_j(f, t)|}
\end{align}
where $X_i(f, t)$ is the STFT of i-th channel, $\mathcal{F}^{-1}$ is inverse Fourier Transform, and $M, F, T$ are the number of microphones, the number of Mel frequency bands, and the number of frames, respectively. Finally, we concatenate $X^{\text{Mel}}$ and $X^{\text{GCC}}$ in the channel dimension and feed the combined tensor to the encoder. In cases of 4 microphones, each patch comprises a total of 10 channels: 4 channels of Mel spectrograms and 6 channels of GCC features.

\subsection{Multi-level Data Augmentation} \label{mlda} Data augmentation plays a pivotal role in contrastive learning and directly influences the quality of the resulting representations \cite{simclr, good_views}. Previous research on single-channel audio \cite{cl_sound, byol_audio} has employed augmentation strategies like \textit{Mixup} \cite{mixup}, \textit{RandomResizeCrop}, \textit{SpecAugment} \cite{specaugment}, \textit{Compression}, and more, specifically on Mel spectrograms. In our Multi-level Data Augmentation pipeline, we apply \textit{Mixup} to multi-channel waveforms instead of Mel spectrograms and \textit{RandomResizeCrop} concurrently on all channels of both Mel and GCC features. Furthermore, we introduce novel channel-wise data augmentation methods \textit{ChannelSwap}, which operates on the order of the microphones, and \textit{ChannelDrop}, which randomly masks Mel and GCC channels.

We describe each data augmentation method in the sequence in which they are applied to the input data, starting from the waveform and progressing through the Mel and GCC features before reaching the encoder. There is flexibility to integrate additional data augmentation in a similar manner.

\subsubsection{ChannelSwap}
\textit{ChannelSwap} rearranges the microphone order of a waveform from a circular or tetrahedral microphone array, resulting in a new recording that maintains the spectral characteristics while altering the spatial location, without requiring additional recording or simulation. In the case of a 4-channel circular array positioned parallel with the ground, there are a total of 8 possible channel rearrangements. Each of them corresponds to the same event with a flipped (multiplied by 1 or -1) and rotated (added by $-90^{\circ}$, $0^{\circ}$, $90^{\circ}$, or $180^{\circ}$) azimuth.

The original intention behind \textit{ChannelSwap} was to boost supervised SELD performance with limited data by generating 8 times more recordings with 8 altered directions \cite{four-stage}.
In our case, although we do not know the azimuth value before or after  \textit{ChannelSwap}, we notice that two samples of similar directions (i.e. a positive pair) prior to \textit{ChannelSwap} should continue to exhibit similar directions after the operation. Thus, we apply \textit{ChannelSwap} with the same but random arrangement for both patches of one utterance to generate more pairs of positive samples.

\subsubsection{Mixup}
\textit{Mixup} mixes the target source $x$ with another random source $y$ as the background, following a convex combination:
\begin{align}
    x^{m} = (1 - \alpha) x + \alpha y
\end{align}
$\alpha$ is a small number closer to 0. Pervious works \cite{cl_sound, byol_audio} mix single-channel Mel spectrograms of two sources. However, we mix multi-channel waveforms instead to fuse both spectral and spatial characteristics at the same time. It is worth noting that we perform waveform normalization both prior to ensure the prominence of the target source and afterward to counteract any statistical shifts.

\subsubsection{RandomResizedCrop}
\textit{RandomResizedCrop} performs resizing and cropping on input images. In the context of spatial audios, both the Mel and GCC features can be regarded as 2D images. Application of \textit{RandomResizedCrop} on the Mel features induces pitch shifting and time stretching effects. In the case of GCC features, besides time stretching, it induces a minor perturbation in the source direction. This strategic adjustment facilitates the generation of positive pairs characterized by closely neighboring directions, all without necessitating precise knowledge of the exact location of these sources.

\subsubsection{ChannelDrop}
Finally, we propose \textit{ChannelDrop} for multi-channel features. \textit{ChannelDrop} in the channel dimension is analogous to \textit{SpecAugment} in the time and frequency dimension. For both Mel and GCC channels, we randomly mask the entire channel to all zeros with a small probability $p$. We observe that, without strong supervision for all the tasks of interest, the model has a tendency to fixate on a single channel or a small subset of channels, leading to a sub-optimal solution for downstream tasks. \textit{ChannelDrop} effectively addresses this potential channel bias by dropping entire channels, compelling the model to attend to all channels and both Mel and GCC features. Consequently, this promotes the learning of a more comprehensive and resilient representation featuring both class and location information.

\subsection{Projection Head and Contrastive Loss} \label{loss}
We adopt the same projection head and contrastive loss as the single-channel counterpart \cite{simclr, cl_sound}. Given a positive pair denoted as $x_i$ and $x_j$, we derive embeddings $h_i$ and $h_j$ after feature extraction, data augmentation and encoder. Subsequently, a two-layer perceptron projects $h_i$ and $h_j$ into $z_i$ and $z_j$, respectively. The contrastive loss, specifically \textit{NT-Xent}, between this pair is defined as follows:
\begin{align}
    \mathcal{L}_{i, j} = -\log \frac{\exp(\text{sim}(z_i, z_j) / \tau)}{\sum_{k=1}^{2N} \mathds{1}_{[k \neq i]} \exp(\text{sim}(z_i, z_k) / \tau)}
\end{align}
where ``sim" is cosine similarity and $\tau = 0.1$ is the temperature. The projector $\mathcal{P}$ is discarded for the supervised evaluation. $h_i$ and $h_j$ are the extracted spatial embeddings.

\section{Experiments}
\label{sec:experiments}

\subsection{Data Simulation}
We simulated 4-channel recordings using sources from FSDnoisy18k dataset \cite{fsdnoisy18k} with gpuRIR toolbox \cite{gpuRIR}. FSDnoisy18k comprises single-channel audio sources, each belonging to one of 20 different classes, such as engines, footsteps, and guitars. For each source, we simulated a room with width, length, and height uniformly sampled from the ranges of [3.0, 10.0], [3.0, 10.0], and [2.5, 4.0] meters, and with a reverberation time (RT60) uniformly sampled from [0.1, 1.0] seconds. We also modeled a circular omnidirectional microphone array with a diameter of 0.1 meters, equipped with 4 evenly spaced microphones. The source and the array were positioned at random locations within the room, with their heights chosen in [0.5, 2.0] meters. To ensure proper placement, they were situated at a minimum distance of 0.5 meters from both the room walls and each other.

FSDnoisy18k contains a larger \textit{noisy} training set (15,813 clips / 38.8 hours), a smaller \textit{clean} training set (1,772 clips / 2.4 hours), and a testing set (947 clips / 1.4 hours). For validation purpose, 200 clips were randomly chosen from \textit{clean}. Each experiment includes two steps: self-supervised pre-training on \textit{noisy} (Section \ref{pre-train}) and supervised evaluation on \textit{clean} (Section \ref{eval}). We reported the model's performance on the testing set.

\subsection{Model and Pre-training} \label{pre-train}
We used a convolutional recurrent neural network (CRNN) adapted from \cite{cl_sound, fsd50k} as our encoder. The embedding dimension is 128. To examine the effect of each data augmentation method in the MC-SimCLR pre-training step, we sequentially introduced \textit{ChannelSwap} (CS), \textit{ChannelDrop} (CD), \textit{Mixup} (MU), and \textit{RandomResizedCrop} (RRC), one at a time. Unless specified otherwise, we set the dropping probablity $p=0.1$ for CD, the mixing coefficient $\alpha$ sampled from [0, 0.01] for MU, and the cropping scale in [0.8, 1] and the aspect ratio in [0.8, 1.25] for RRC. Table \ref{table:cd} shows the variation in performance with different dropping probability for CD.

For all pre-training experiments, we used a batch size of 512 and a SGD optimizer with a learning rate of 0.2, a momentum value of 0.9, and a weight decay factor of 0.0001. We trained the encoder for 500 epochs, initiating with a linear learning rate warmup for the first 10 epochs and succeeding a cosine learning rate decay \cite{sgdr} to 0 for the remaining epochs. All experiments were conducted on an NVIDIA L40 GPU.

\begin{table}[htb]
	\centering
	\caption{Linear-probing and fine-tuning results with different combination of data augmentation methods. Our best model (colored yellow) significantly outperforms the baseline model without pre-training (colored gray) in both tasks with both evaluation protocals.\\} 
    \begin{adjustbox}{width=\columnwidth,center}
    \begin{tabular}{>{\raggedright}p{5cm}|cc|cc}
    \hline
    \hspace{0.4cm} Evaluation & \multicolumn{2}{c}{Linear-probing}  & \multicolumn{2}{c}{Fine-tuning} \\
     \diagbox[innerwidth=5cm, innerleftsep=20pt, innerrightsep=20pt]{Pre-training}{\rotatebox{-10}{Metrics}} & Accuracy$\%$ & Error$^{\circ}$ & Accuracy$\%$ & Error$^{\circ}$ \\
     \hline
     \rowcolor{gray!20} \hspace{0.35cm} \hspace{0.3cm} \textit{Random} &  23.6 &  83.1 & 43.7 & 11.8 \\
    \hline
    \hspace{0.75cm} \textit{MC-SimCLR} w/o DA  &  33.0 &  13.2 & 45.6 & 11.4 \\
    \hspace{1cm} + CS  &  34.5 &  12.2 & 46.1 & 10.6 \\
    \hspace{1cm} + CS + CD &  43.6 &  11.3 & 50.9 & 9.5 \\
    \hspace{1cm} + CS + MU + CD &  47.8 &  10.7 & 51.1 & 9.3 \\
    \hspace{1cm} + CS + RRC + CD &  49.8 &  9.4 & 52.8 & 8.6 \\
    \rowcolor{yellow!20} \hspace{1cm} + CS + MU + RRC + CD &  51.5 &  10.1 & 53.4 & 8.7 \\
    \hline
    \end{tabular}
    \end{adjustbox}
    \label{table:final}
\end{table}

\begin{figure}[htb]
  \centering  \includegraphics[width=\columnwidth]{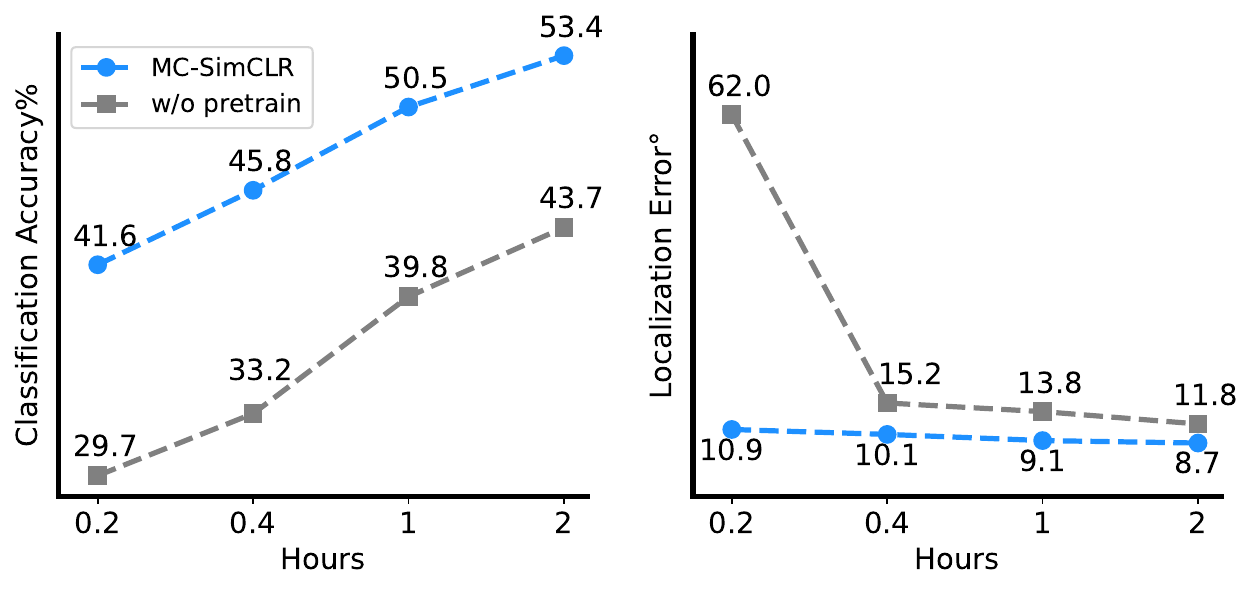}
  \caption{Fine-tuning with different amounts of labeled data. The performance gap between pre-trained and random-initialized models is more significant when labeled data are scarce for fine-tuning.}
  \label{fig:subset}
\end{figure}


\subsection{Supervised Evaluation} \label{eval}
After pre-training, we evaluated the SELD performance of the representation using the following three protocols:
\begin{itemize}
    \item \textbf{Linear-probing} We individually trained two linear layers for event classification and azimuth prediction with the encoder's weights fixed.
    \item \textbf{Fine-tuning} We jointly trained two linear layers and simultaneously update the encoder weights.
    \item \textbf{Subset Fine-tuning} Same as Fine-tuning, but we only use a subset of the labeled data.
\end{itemize}
We followed the same hyperparamters as the pre-training step, with the exception of using a smaller batch size of 128 and a learning rate of 0.01. No data augmentation is used except $\textit{ChannelSwap}$, as it is proved to boost supervised SELD performance \cite{four-stage}. Additionally, if fine-tuning a pre-trained model, we applied a scaling factor of 0.1 to the encoder's gradients to prevent overfitting on \textit{clean}.

\section{Results}
\label{sec:results}

In Table \ref{table:final}, we provide a comparative analysis of event classification accuracy and azimuth prediction error across combinations of data augmentation methods. The first row represents the supervised baseline without any pre-training: A random encoder without pre-training erases nearly all spatial information, as shown from the linear-probing error, which is close to the level of guessing (90$^{\circ}$). While all pre-trained models outperform the baseline in both tasks, the inclusion of data augmentation further enhances performance. There is a progressive performance improvement with the addition of each augmentation method, except a small increase in error with \textit{Mixup} on top of \textit{ChannalDrop}. Notably in linear-probing, the inclusion of \textit{ChannalDrop} results in an accuracy boost of around 9\%, and the inclusion of \textit{RandomResizeCrop}  leads to an improvement of more than 6\% in accuracy and a reduction of around 2$^{\circ}$ in error. Our best result is achieved when we incorporate all four augmentation methods in the final row: this results in a 7.8\% improvement in accuracy and a 1.7$^{\circ}$ reduction in error when fitting only the last linear layers, compared to end-to-end training without pre-training. Fine-tuning the pre-trained model yields an even more substantial improvement, with a 9.7\% increase in accuracy and a 3.1$^{\circ}$ reduction in error.

The advantages of MC-SimCLR pre-training become even more pronounced with limited labeled data. To demonstrate this, we reduced the amount of labeled data available for fine-tuning and compared the resulting performance for models with or without pre-training in Fig \ref{fig:subset}. We notice that a randomly initialized model fails to learn localization with only 0.2 hours of data, and the classification accuracy lags behind that of a pre-trained model by more than 10\%. In contrast, the pre-trained model showcases its ability to achieve comparable or superior performance with substantially smaller datasets: In comparison to training from scratch using the entire dataset, a pre-trained model achieves comparable performance (2.1\% lower accuracy but 0.9$^{\circ}$ lower error)
with only 0.2 hours and exhibits superior performance in both tasks (2.1\% higher accuracy and 1.7$^{\circ}$ lower error) with only 0.4 hours of the data.

\begin{table}[t]
    \centering
	\caption{Fine-tuning performance vs \textit{ChannelDrop} probablity $p$. This first row in gray corresponds to no \textit{ChannelDrop}. A modest $p$ benefits both the classification and localization tasks. However, a high $p$ (meaning excessive feature dropping) has a detrimental impact on localization error. We choose $p=0.1$ (colored yellow) in this paper.
 \\} 
    \begin{adjustbox}{width=0.4\columnwidth}
    \begin{tabular}{c|cc}
    \hline
    $p$ & Accuracy$\%$ & Error$^{\circ}$ \\
    \hline
    \rowcolor{gray!20} 0  &  51.3 &  14.4 \\
    0.05  &  53.5 &  8.8 \\
    \rowcolor{yellow!20} 0.1  &  53.4 &  8.7 \\
    0.2  & 54.2 &  22.1 \\
    \hline
    \end{tabular}
    \end{adjustbox}
    \label{table:cd}
\end{table}

\section{Conclusion}
\label{sec:conclusion}
This study introduces MC-SimCLR, a specialized contrastive learning framework for multi-channel audio signals. MC-SimCLR adeptly leverages unlabeled spatial audio recordings to extract a strong combined spectral and spatial representation. This representation is enhanced through a multi-level multi-dimensional data augmentation pipeline spanning waveforms, Mel spectrograms, and GCC features during training. Empowered by these augmentations, MC-SimCLR results in substantial improvements in event classification and localization compared to training from scratch. Our future work will be directed toward learning the representation of moving or overlapping spatial sound events.

\section{Acknowledgement}
This work was funded by the National Institutes of Health (NIH-NIDCD) and a grant from Marie-Josee and Henry R. Kravis.

\bibliographystyle{IEEEbib}
{\footnotesize \bibliography{strings,refs}}

\end{document}